# Bulk-scale synthesis of randomly stacked graphene with high crystallinity

Zizhao Xu[a*], Shingo Nakamura[a], Taiki Inoue[a], Yuta Nishina[b], Yoshihiro Kobayashi[a*]

[a] Department of Applied Physics, Graduate School of Engineering, Osaka University, 2-1 Yamadaoka, Suita, Osaka, 565-0871, Japan

[b] Research Core for Interdisciplinary Sciences, Okayama University, 1-1-1 Tsushima-naka, Kita-ku, Okayama, 700-8530, Japan

*Corresponding authors.

Email addresses: zizhao_xu@ap.eng.osaka-u.ac.jp (Zizhao Xu), kobayashi@ap.eng.osaka-u.ac.jp (Yoshihiro Kobayashi).

**Abstract**

Since the strong interlayer interaction of AB-stacked graphene in bulk form degrades the superior property of single-layer graphene, formation of randomly stacked graphene is required to apply the high performances of graphene to macroscopic devices. However, conventional methods to obtain bulk-scale graphene suffer from a low crystallinity and/or the formation of a thermodynamically stable AB-stacked structure. This study develops a novel approach to produce bulk-scale graphene with a high crystallinity and high fractions of random stacking by utilizing the porous morphology of a graphene oxide sponge and an ultrahigh temperature treatment of 1500–1800 °C with ethanol vapor. Raman spectroscopy indicates that the obtained bulk-scale graphene sponge possesses a high crystallinity and a high fraction of random stacking of 80%. The large difference in the random-stacking ratio between the sponge and the aggregate samples confirms the importance of accessibility of ethanol-derived species into the internal area. By investigating the effect of treatment temperature, a higher random-stacking ratio is obtained at 1500 °C. Moreover, the AB-stacking fraction was reduced to less than 10% by introducing cellulose nanofiber as a spacer to prevent direct stacking of graphene. The proposed method is effective for large-scale production of high-performance bulk-scale graphene.

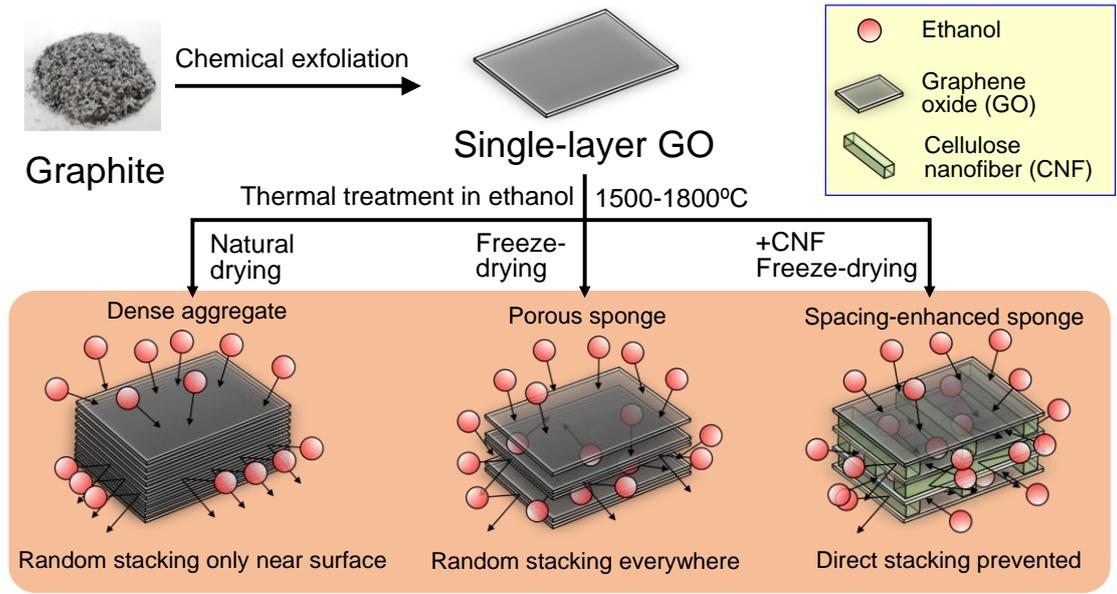

Keywords: Bulk-scale graphene; Reduced graphene oxide; Thermal reduction; Stacking order; Interlayer interaction

# 1. Introduction

Graphene possesses numerous excellent properties such as a high carrier mobility, electrochemical performance[1], optical transparency, thermal conductivity[2], and mechanical strength[3]. These properties are attributed to the unique electronic structure derived from a one-atom-thick honeycomb lattice of single-layer graphene. Due to these excellent properties, graphene and graphene-containing materials have been studied extensively toward applications in electronics[4-6], electrode material[7-11], etc. One problem is that the thinness and small volume of a single-layer graphene flake limits the electrical, mechanical, and other performances. Thus, bulk-scale graphene, which is an aggregate composed of plenty of graphene flakes, is required for various daily applications[12] such as pressure sensors[13, 14] and battery electrodes[7-11]. Preparation of high-quality bulk-scale graphene is a critical issue for practical applications.

The production of bulk-scale graphene starting from graphene oxide (GO) is a promising approach due to mass-production compatibility and structure controllability. In this production process, GO flakes are dispersed in solution through the functionalization of bulk graphite with oxygen-containing groups. GO then reduced into reduced graphene oxide (rGO). The most common approach for reduction is a hydro-thermal method or a chemical method [15, 16], but these do not address the defect issue, such as vacancy. Alternatively, a high-temperature treatment method can produce rGO with the highest crystallinity[17]. However, the stacking structure of multilayer graphene is problematic. The thermodynamically favorable AB-stacked structure of multilayer graphene is formed during reduction of GO at high-temperature. A strong interlayer interaction in AB-stacked multilayer graphene causes its electronic structure to deviate from that of single-layer graphene, degrading the superior properties[18]. On the other hand, theoretical calculations have predicted that randomly stacked graphene, where adjacent graphene layers are randomly rotated or translated, can preserve the properties of single-layer graphene because it has an electronic structure similar to that of single-layer graphene [19]. Experimental studies have confirmed that randomly stacked graphene keeps a single-layer-like electronic structure[20] and has superior properties compared to AB-stacked graphene. Richter et al.[21] realized a high mobility of $7 \times 10^4$ cm$^2$/V·s for individual flakes of multilayer graphene with a rotationally stacked structure, while Liu et al.[18] found that AB-stacked bilayer graphene films exhibited a mobility of $4.4 \times 10^3$ cm$^2$/V·s. The development of fabrication methods for bulk-scale graphene with controlled interlayer stacking is crucial to realize graphene-based applications in numerous fields.

We previously produced graphene with a high fraction of the randomly stacked structure from GO aggregates by ultrahigh temperature reduction under an ethanol vapor supply[22]. At ultrahigh temperature, ethanol is decomposed into different species and further chemical reactions in gas phase occurs [23]. Reaction products containing carbon atoms and OH groups (hereafter called as "ethanol-derived species") mainly act as carbon source and etchant, respectively. This reduction method has also been utilized for few-layer rGO on substrates and achieved a high carrier mobility[24]. However, the analysis of the bulk-scale graphene was limited to the outer surface of a GO aggregate[22], and the stacking structure of the internal area was not clarified. The repairing process and the formation of a randomly stacked structure should be limited to the surface area of the GO aggregates because the ethanol-derived species cannot enter the internal area of dense samples.

Accessibility of ethanol-derived species to GO flakes should be a critical factor for the successful formation of a randomly stacked structure induced by ethanol-mediated reduction of GO. Preparation of a GO sponge with a highly porous structure by freeze-drying is a potential solution for the inaccessibility problem of GO flakes[25]. Freeze-drying is a method by which liquid-containing materials are frozen below the freezing point and the solvent is sublimated into a vapor and removed under vacuum. The original structure and shape of the material are maintained because the solvent in the solid phase is sublimated directly into gas phase. Instead of aggregation, the GO dispersion can be dried into a GO sponge where GO flakes maintain separation similar to the dispersion. It should be mentioned that GO flakes randomly rotated in three dimensions, which was less impacted by nearby GO flakes compares to aggregate or film samples. The as-prepared GO sponge has a porous structure with a large surface area, which is suitable for further reduction reactions.

Besides inhibiting AB-stacked structure formation during the reduction process, the addition of other materials as spacers may effectively prevent graphene layers from stacking physically. Spacers must be chemically inert or transformed into inert materials at high temperature to prevent reactions between GO and spacers. Additionally, the spacers must be water soluble to promote molecular level mixing of GO and the spacers. Cellulose nanofiber (CNF) fulfills these requirements for spacers. CNF is a natural cylindrical polymer with plenty of resources, renewability[26], high strength, high stiffness[27-30], and a low weight[14, 31], which makes it widely used in many areas[32-34]. CNF can be prepared by (2,2,6,6-tetramethylpiperidin-1-yl)oxidanyl, which is known as the TEMPO method. CNF possesses a high aspect ratio of 4–10 nm in diameter and 1 μm in length[26], which is suitable for intercalation into the GO interlayers in a dispersion before reduction. The water-solubility was improved by carboxylate groups on CNF [35, 36]. These effects make CNF a promising candidate as a spacer for graphene.

Herein we propose a method that combines freeze-drying and an ultrahigh temperature process to produce bulk-scale graphene to tackle property degradation due to the strong interlayer interaction in AB-stacked structure. This method can repair and reduce GO on the bulk scale and realize a high random-stacking fraction, which should preserve the properties of single-layer graphene. A GO sponge prepared by freeze-drying of a GO dispersion presents a larger surface area than a GO aggregate. An increase in the accessible area of the GO sponge by ethanol vapor contributes to a high crystallinity and high fractions of random stacking. Additionally, this study provides an approach to further decrease the AB-stacked structure ratio of a graphene sponge utilizing CNF. CNF serves as a spacer that intercalates between the graphene layers to prevent the stacking where graphene layer contact directly (direct stacking). This rGO/CNF sponge features a low defect density, large surface area, reduced interlayer stacking, and bulk-scale production compatibility, increasing its potential applicability.

## 2. Materials and methods

*2.1 Preparation of rGO sponge*

Figure 1 schematically illustrates the fabrication process of bulk-scale graphene samples. GO was prepared from graphite by a modified Hummers' method[37]. The obtained GO dispersion was 1 wt% in water solvent, and the flake size of GO was about 10 μm (observed by optical microscopy). The following freeze-drying process was carried out by a lab-made vacuum drying system to prepare a GO sponge. A GO dispersion was added into an ice tray and shaped into a 1-cm$^3$ cube. It was then frozen in a freezer at −10 °C overnight followed by pumping for 48 hours. Pumping sublimated the water in the frozen GO dispersion, leaving the GO network structure as a GO sponge. Then the GO sponge was thermally treated in ethanol/Ar gas under ultrahigh temperature conditions for repair and reduction. Instead of the solar furnace used in our previous study[22], the ultrahigh temperature process was performed at 1800 or 1650 °C using an infrared radiation furnace (SR1800G-S, THERMO RIKO Co.) or at 1500 °C using a tubular electric furnace (FT-01VAC-1650, FULL-TECH Co.). Both of the furnaces were connected to vacuum pumps to maintain a low pressure. We introduced 20 sccm of Ar under the total pressure of 26.6 Pa during temperature rise. After reaching the set temperature, the thermal treatment was conducted at ultrahigh temperature with flowing 100 sccm of Ar and 0.3 sccm of ethanol under the pressure of 106.6 Pa. The obtained samples, which were named GS-Et1800, GS-Et1650, and GS-Et1500, respectively, were cut by a cutter to characterize the stacking structure of the internal area by Raman spectroscopy. The surface of GS-Et1800, named GS-Et1800-surf, was also characterized by Raman spectroscopy for comparison with the internal area. GS-Ar1800 was prepared by the same procedure at 1800 c without ethanol vapor (100 sccm of Ar under the total pressure of 106.6 Pa), and GS-Ar1800-surf was readied for Raman spectroscopy. Measurement condition of Raman spectroscopy will be stated in the following section.

*2.2 Preparation of rGO aggregate*

For comparison with the porous GO sponge samples, GO aggregates were prepared by naturally drying a GO dispersion and a subsequent thermal treatment for repair and reduction in an infrared radiation furnace in Ar gas or ethanol/Ar gas at 1800 °C[22], named GA-Ar1800 and GA-Et1800, respectively. The inside of GA-Et1800 was examined by Raman spectroscopy after cleaving with adhesive tape to remove the surface part. The surface parts of the aggregate samples were also characterized (GA-Ar1800-surf and GA-Et1800-surf).

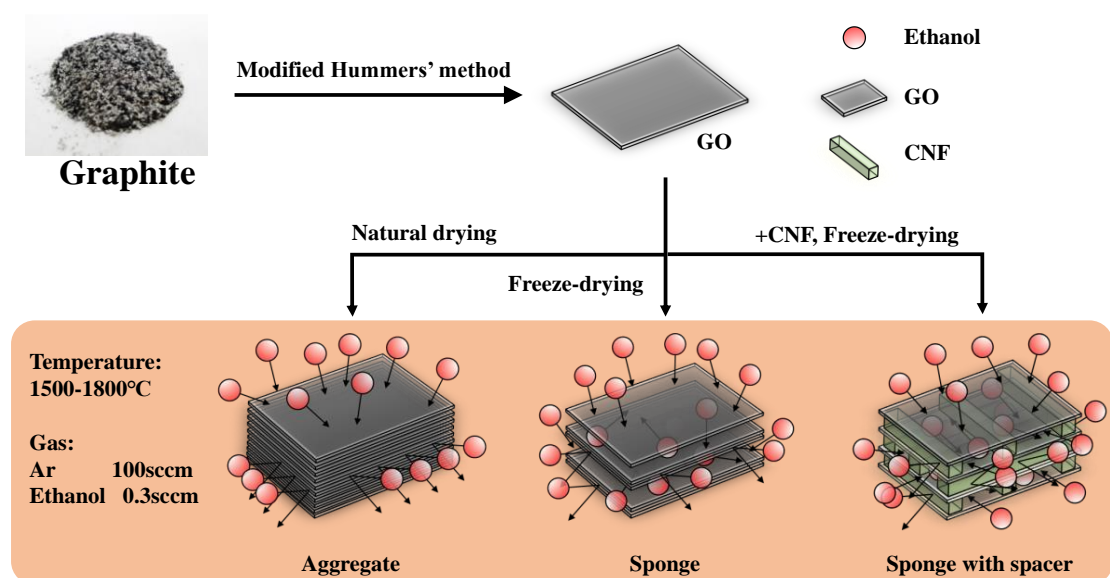

Fig. 1. Schematic illustrations of the bulk-scale graphene fabrication process. GO aggregate, GO sponge, and GO/CNF sponge are prepared from a GO dispersion by natural drying, freeze-drying, and freeze-drying with the addition of CNF, respectively. GO samples with different morphologies are then reduced into rGO by ethanol vapor associated reduction under ultrahigh temperature conditions.

*2.3 Preparation of a composite sponge of reduced graphene oxide and cellulose nanofiber*

CNF was prepared by the TEMPO method and used as received from DKS Co.[26]. It featured a high aspect ratio of 4–10 nm in diameter and 1 μm in length. The GO dispersion was blended with CNF in a mixer for 2 min at 10000 rpm and frozen in a freezer at −10 °C overnight followed by pumping for 2 days. The total mass fraction of the dispersion was 1 wt% with a mass ratio of GO: CNF = 4:6. The obtained GO/CNF sponges were thermally treated for repair and reduction at 1500 °C in ethanol/Ar gas. This composite sponge was named GCS-Et1500.

A CNF sponge without GO was also prepared for comparison. CNF was blended with water in a mixer for 2 min at 10000 rpm and frozen in a freezer at −10 °C overnight followed by pumping for 2 days. The total mass fraction of dispersion was 1 wt%. The obtained CNF sponges were thermally treated for repair and reduction at 1500 °C in ethanol/Ar gas. This CNF sponge was named CS-Et1500.

*2.4 Characterization*

Raman spectra were obtained by LabRAM HR-800 UV (Horiba Jobin Yvon) with an excitation laser wavelength of 532 nm, power of 1 mW, and spot size of 1 μm. The Raman spectra of GA-Et1800, GA-Et1800-surf, GS-Ar1800-surf, GS-Ar1800, GS-Et1800-surf, GS-Et1800, and GS-Et1650 were obtained by averaging spectra measured at five random spots. Raman spectra of GS-Et1500 and GCS-Et1500 were averaged for 100 random spots. The fraction of the randomly stacked structure was calculated by G′-band fitting since the G′-band is sensitive to structural changes that are vertical to the graphene plane[38, 39]. The G′-band profile was composed of several peaks originating from both randomly stacked and AB-stacked structures[38]. Note that G'-band is also called 2D-band in the literature[40]. For distinction from two-dimensional (2D), we use the notation G'-band in this paper. Details of the fitting process are described in the next section. Scanning electron microscopy (SEM) images were taken by VE-8800 (Keyence) at an acceleration voltage of 15 kV.

**3. Results and discussion**

*3.1 Effect of accessibility of ethanol-derived species on the random-stacking fraction*

*of graphene*

The Raman spectra of the as-prepared rGO sponge and aggregate samples were measured with an exposure time and accumulation of 15 s and 10 times, respectively (Fig. 2(a)). The G-band and D-band of these samples were observed around 1580 cm$^{-1}$ and 1350 cm$^{-1}$, respectively. The former derives from in-plane stretching mode of the hexagonal lattice of graphene, and the latter originates from hexagon-breathing mode activated through the presence of lattice defects[41]. The intensity ratio of the D-band to the G-band, $I_D/I_G$, which corresponds to the defect density of graphene, ranged 0.09–0.37. Note that the relationship between D-band intensity and defect density is classified into two different stages depending on defect density, namely interdefect distance. In the case of a small interdefect distance less than ~3 nm D-band intensity decreases with increasing defect density, named as stage 1, whereas in the case of large interdefect distance larger than ~3 nm D-band intensity increases with increasing defect density, named as stage 2[42]. The stage 1 and stage 2 can be distinguished by the broadening of D-band. The full width at half maximum of D-band of our samples ranged 35–59 cm$^{-1}$, indicating that they were on the stage 2 [43, 44]. The narrow D-band shapes and the low $I_D/I_G$ ratios confirm that the present graphene samples have much lower defect density than rGO obtained by chemical reduction[42] or annealing in ethanol at ~1000 °C[45-47]. It should be noted that the rGO formed by chemical reduction dose not proceed to stage 2 but remains on stage 1. The result indicates the ultrahigh temperature process is effective to produce highly crystalline graphene. SEM images showed a dense and flat structure of the aggregate sample, while a porous structure was observed on the thermo-treated rGO sponge (Figs. 2 (b) and (c)), confirming that freeze-drying successfully forms a high surface area structure. Note that the D band originates from both edge area and point defect[48], which will be analyzed in details in a future work.

The $I_D/I_G$ ratios for the surface area of ethanol-treated aggregate samples (GA-Et1800-surf) and sponge samples (GS-Et1800-surf) indicated similarly low defect densities ($I_D/I_G$ of 0.09 and 0.14, respectively (Fig. 2 (a)). The internal area of the aggregate sample and sponge sample under the ethanol condition (GA-Et1800 and GS-Et1800, 0.32 and 0.36), and the surface area of aggregate samples and sponge samples under the Ar gas condition (GA-Ar1800-surf and GS-Ar1800-surf, 0.37 and 0.26) showed relatively high $I_D/I_G$ ratios compared with the surface of the samples treated under ethanol condition (GA-Et1800-surf and GS-Et1800-surf). These results indicated that the behavior of defect healing depends mainly on the accessibility of ethanol-derived species to GO instead of its morphology.

The G′-band of the graphene samples was observed around 2700 cm$^{-1}$ with the different sharpness and intensity (Fig. 2 (a)). The intensity ratio of the G′-band to the G-band, $I_{G'}/I_G$, provides information about the interlayer interactions of graphene because the G′-band is sensitive to the layer number and stacking order of graphene[49-51]. Higher $I_{G'}/I_G$ ratios were observed in GA-Et1800-surf, GS-Et1800-surf, and GS-Et1800, indicating a smaller interaction between graphene layers. GA-Et1800 and GA-Ar1800-surf displayed low $I_{G'}/I_G$ ratios, suggesting a stronger coupling between adjacent graphene layers, while GS-Ar1800-surf showed a moderate $I_{G'}/I_G$ ratio. This was attributed to the fact that rGO layers have a stronger tendency to form an AB-stacked structure in the aggregate shape, even crumples were located on rGO layers. On the other hand, a sponge shape provides a high fraction of the randomly stacked structure with a weak interaction between the graphene layers.

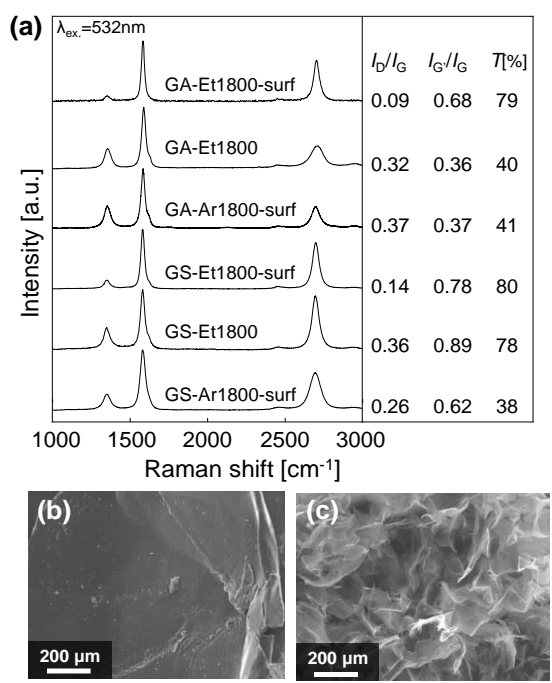

Fig. 2. (a) Raman spectra of GA-Et1800-surf, GA-Et1800, GA-Ar1800-surf, GS-Et1800-surf, GS-Et1800, and GS-Ar1800-surf. D-band, G-band, and G′-band are observed at ~1350 cm$^{-1}$, 1580 cm$^{-1}$, and 2700 cm$^{-1}$, respectively. Intensity ratios of D-band to G-band, $I_D/I_G$, G′-band to G-band, $I_{G'}/I_G$, and random-stacking ratio, $T$, obtained from the G′-band analysis are displayed to the right of the corresponding spectra. (b, c) SEM images of (b) GA-Et1800-surf and (c) GS-Et1800.

To evaluate the stacking order of graphene in the bulk-scale samples, we further analyzed the G′-band peak shape. According to Cançado et al.[38], the G′-band in the Raman spectrum of multilayer graphene can be fitted by three Lorentzian peaks. The first peak was located around 2700 cm$^{-1}$, which is close to the frequency of the original G′-band of single layer graphene. It is denoted as a G′$_{2D}$ component because it is associated with a two-dimensional graphite feature in which the stacking order along the c-axis is low, namely a turbostratic structure or randomly stacked structure. The two other peaks around 2680 cm$^{-1}$ and 2720 cm$^{-1}$ are denoted as G′$_{3DA}$ and G′$_{3DB}$ components, respectively. These are related to the three-dimensional configuration of graphite, that is, the AB-stacked structure. It should be mentioned that intensity of G'$_{3DB}$ is proportional to the volume of 3D graphitic regions [38, 52]. We employed Cançado's method to calculate the random-stacking fraction in the bulk-scale graphene.

From the intensities of the three components, the random stacking-fraction of graphene, *T*, can be written as

$$T\ [\%] = \frac{I_{G'_{2D}}}{I_{G'_{3DB}} + I_{G'_{2D}}} \times 100 \qquad (1)$$

where $I_{G'_{2D}}$ and $I_{G'_{3DB}}$ denote the intensities of G′$_{2D}$ and G′$_{3DB}$ peaks, respectively. For an ideal random-stacking structure of multilayer graphene, the Raman spectrum shows single-layer graphene-like features, *i.e.*, a strong G′$_{2D}$ peak and negligible G′$_{3DA}$ and G′$_{3DB}$ peaks, which will result in a *T* value close to 100%. It should mention that intensity of G′$_{3DB}$ is for volume of graphitic regions, while G′$_{2D}$

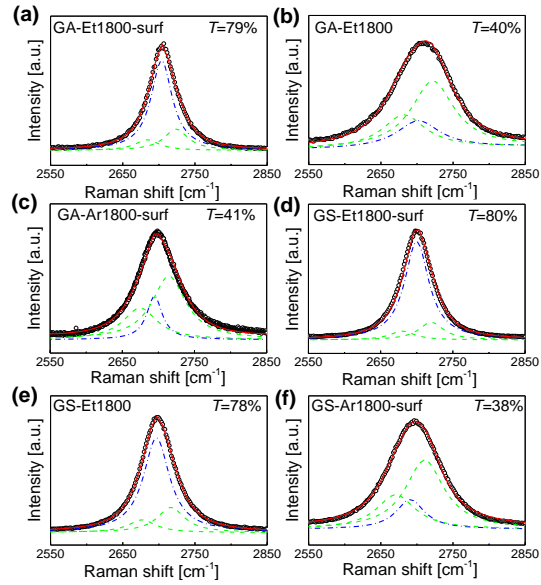

Fig. 3. G′-band fitting of (a) GA-Et1800-surf, (b) GA-Et1800, (c) GA-Et1800-surf, (d) GS-Et1800-surf, (e) GS-Et1800, and (f) GS-Ar1800-surf, which provides randomly stacked structure ratio, *T*. Black open circles denote the measured spectra. Blue dashed-

dotted curves denote the $G'_{2DA}$ components originating from the randomly stacked structure of graphene, while the two green dashed curves denote the $G'_{3DA}$ and $G'_{3DB}$ components derived from the AB-stacked structure. Red solid curves are sum of the blue and green curves.

Via stacking order analysis based on the $G'$-band[38], as shown in Fig. 3 (a–e), GA-Et1800-surf, GS-Et1800-surf, and GS-Et1800 displayed high $T$ value of 79%, 80%, and 78%, respectively. By contrast, GA-Et1800 showed a low $T$ value of 40%, that is, it had a high AB-stacking fraction. Although AB-stacking is the thermodynamically stable structure, the superior properties originating from single-layer graphene are degraded. The high random-stacking fractions of GA-Et1800-surf, GS-Et1800-surf, and GS-Et1800 indicated that the formation of AB-stacked structure was effectively suppressed. As shown in Fig. 2 (b), the rGO aggregate possessed a dense structure, making the internal area inaccessible for the ethanol-derived species. Relatively low $T$ indicates preferential formation of AB stacking structure under Ar environment, even rGO layers were deformed by crumple naturally induced for stacked GO layers. The apparent difference between the surface and internal area of the aggregate samples (GA-Et1800-surf and GA-Et1800, Figs. 3 (a) and (b)) indicated that, as expected, the formation of a randomly stacked structure is restrained on the surface and does not occur in the internal area of the aggregate. It should be mentioned that intensity of $G'_{3DB}$ is directly proportional to the volume of 3D graphitic regions, and the overall intensity from 2D and 3D graphitic regions is proportional to ($G'_{3DB}$ + $G'_{2D}$) since the intensity ratio of $G'_{3DB}$ to $G'_{3DA}$ is constant ($I_{G'3DB}/I_{G'3DA}$ ~2)[38, 52]. On the other hand, both the surface and internal areas of the sponge samples (GS-Et1800-surf and GS-Et1800, Figs. 3 (d) and (e)) showed similarly high fractions of the randomly stacked structure. These results confirmed that sponge structure provides a higher exposed area of GO flakes for ethanol-derived species, allowing the randomly stacked structure to form even on the inside. For comparison between GA-Et1800 and GS-Et1800, the surface areas (Figs. 3 (a) and (d)) showed a similar result while the internal area of the sponge sample (GS-Et1800, Fig. 3 (e)) had a higher random-stacking fraction under the same reaction conditions. It should be noted that the porous morphology of the GO sponge was not a sufficient factor to obtain high random-stacking fractions. GA-Ar1800-surf and GS-Ar1800 (Figs. 3 (a) and (d)), which were processed with only Ar instead of ethanol/Ar, showed low random-stacking fractions, 34% and 29%, respectively. Thus, the remarkable result of GS-Et1800 was attributed to the effect of ethanol-derived species in the high temperature treatment as well as the accessible surface obtained by the porous structure of GO sponge. Cooperation between the porous structure, ethanol-derived species, and ultrahigh temperature contributed to the formation of bulk-scale graphene with a high crystallinity and high random-stacking fraction. The formation mechanism of randomly stacked graphene by the ethanol-associated ultrahigh temperature process is discussed in the Supporting Information.

*3.2 Effect of process temperature on the random-stacking fraction of graphene*

Multilayer graphene tended to form an AB-stacked structure under a higher temperature process, leading to a higher interlayer interaction and lower G′-band[38, 53]. We tried to decrease the AB-stacking fraction by restraining the movement of the graphene flakes. Reducing the process temperature to 1650 °C or 1500 °C (GS-Et1650 and GS-Et1500) was attempted to increase the fraction of random stacking. Raman spectra of GS-Et1500 were measured with 5 s of exposure time and 5 times for accumulation. The $I_D/I_G$ ratios of GS-Et1650 and GS-Et1500 were 0.18 and 0.53, while the $I_{G'}/I_G$ ratios were 0.77 and 1.2, respectively (Fig. 4 (a)). The low $I_D/I_G$ ratio of GS-Et1650 displayed a similar low defect density as the 1800 °C sample (GS-Et1800-surf). By contrast, GS-Et1500 showed a high $I_D/I_G$ ratio, indicating the relatively low crystallinity. The $I_{G'}/I_G$ ratio of GS-Et1650 was close to that of 1800 °C sample (GS-Et1800), suggesting that GS-Et1650 had a similar interlayer interaction with GS-Et1800. The $I_{G'}/I_G$ ratio of GS-Et1500 was higher than those of GS-Et1650 and GS-Et1800, revealing less coupling between the graphene layers. The $I_{G'}/I_G$ and $I_D/I_G$ exhibit a certain fluctuation for each measurement point as shown in Fig. S2 in Supporting Information. It should be noted that the highest $I_{G'}/I_G$ ratio of 2.3 was observed at a measurement spot of GS-Et1500 (Fig. 4(a), bottom). The SEM image confirmed that the sponge maintained a porous structure after 1500 °C treatment (Fig. 4 (b)), which is similar to that of GS-Et1800 (Fig. 2 (c)). Analysis of the G′-band of GS-Et1650, GS-Et1500, and GS-Et1500 with the highest $I_{G'}/I_G$ ratio showed that the random-stacking fractions were 76%, 80%, and 85%, respectively (Figs. 4 (c–e)). The random-stacking fraction increased as the reaction temperature decreased. The high random-stacking fraction and the highest G′ band of GS-Et1500 suggested that the sample under the reaction condition was far from crossing the energy barrier for AB-stacked structure and preserved a more randomly stacked structure at 1500 °C.

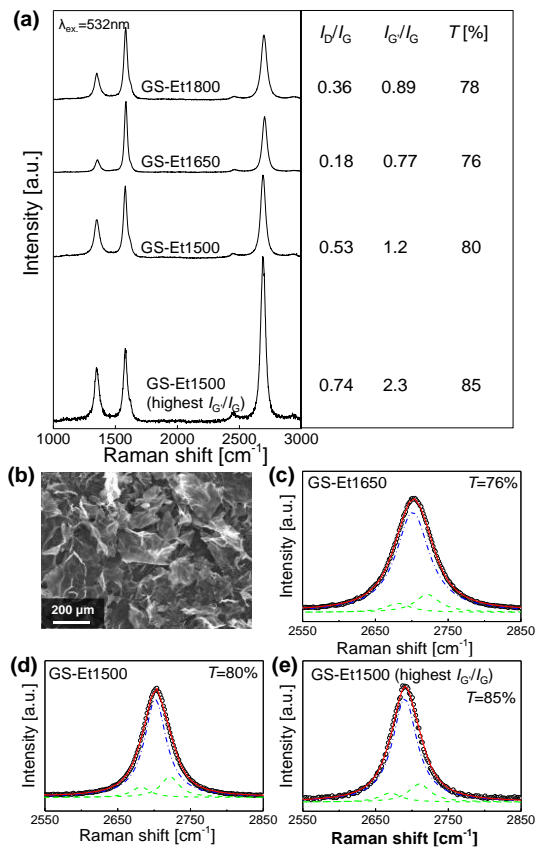

Fig. 4. (a) Raman spectra of GS-Et1800, GS-Et1650, GS-Et1500, and GS-Et1500 with the highest $I_{G'}/I_G$ ratio. Characteristic ratios are indicated to the right of the corresponding spectra. (b) SEM image of GS-Et1500. (c–e) G′-band fitting of (c) GS-Et1650, (d) GS-Et1500, and (e) GS-Et1500 with the highest $I_{G'}/I_G$ ratio. Random-stacking fraction, $T$, is denoted in the graphs.

*3.3 Increase of the random-stacking fraction of graphene by the addition of CNF as a spacer*

We investigated the impact of a spacer to physically prevent graphene from stacking on the interlayer interaction in bulk-scale graphene. Because a high $I_{G'}/I_G$ was achieved in GS-Et1500, the same treatment condition was employed in this experiment. Based on the advantages of CNF, it was added as a spacer into bulk-scale graphene (GCS-Et1500). The Raman spectra of GCS-Et1500 were measured with 5 s of exposure time and 100 times for accumulation. A pure CNF sponge (CS-Et1500) was prepared as the control using the same mixing, freeze-drying, and thermal treatment conditions. The Raman spectra of CS-Et1500 were measured with 5 s of exposure time and 30 times for accumulation. As shown in Fig. 5 (a), the $I_D/I_G$ ratio of GCS-Et1500 was 0.60, while the $I_{G'}/I_G$ ratio was 0.86. CS-Et1500 showed 1.32 of $I_D/I_G$ and 0.26 of $I_{G'}/I_G$. The SEM image of GCS-Et1500 suggested that the sponge also maintained a porous structure (Fig. 5 (b)), which was very similar to the sponge samples treated in ethanol, such as that of GS-Et1800 (Fig. 2 (b)). The CNF structure was not observed by SEM because it was altered through graphitization during the ultrahigh temperature process. By adding CNF as a spacer, GCS-Et1500 showed a higher $I_D/I_G$ and a lower $I_{G'}/I_G$ than the sample without CNF (GS-Et1500). This high D-band intensity was attributed to the graphitization of CNF and formation of amorphous carbon. The G′-band of GCS-Et1500 was analyzed to investigate the stacking order of the composite sponge of graphene and CNF by the analysis procedure described in section 3.2[38]. As shown in Fig. 5 (c), the obtained $T$ value was 93%. It should be noted that the $T$ value here does not simply correspond to the fraction of the randomly stacked structure but indicates the fraction of two-dimensional graphene, which includes randomly stacked graphene and non-stacked graphene separated by the spacer. The high occupation of the two-dimensional graphite-originating peak for GCS-Et1500 indicated that the addition of CNF further suppressed the formation of the AB-stacked structure. It should be emphasized that the $T$ value of GCS-Et1500, 93%, significantly surpassed that of GS-Et1500, 80%. This result implied that inserting CNF physically prevented the direct stacking of the graphene layers.

The spatial variation of the GCS-Et1500 sample was investigated. The $I_{G'}/I_G$ and $I_D/I_G$ ratios of each measurement spot of GCS-Et1500 are plotted in Fig. 5 (d). Most of the Raman results were located at the grey oval circle (region 1) but some Raman results of GCS-Et1500 with a high $I_{G'}/I_G$ (region 2) or a low $I_D/I_G$ (region 3) were observed outside region 1. As schematically illustrated in Fig. 5 (e), this result can be explained as follows. As for region 2 in Fig. 5 (d), CNF was perfectly dispersed and played the spacer role well. After ultrahigh temperature process, CNF was changed into graphite material[54] and remained between graphene layers. Graphene layers were physical separated by the graphitic material, so that interlayer interactions were suppressed. Thus, $I_{G'}/I_G$ was high. However, region 1 showed lower $I_{G'}/I_G$ since CNF was insufficient in these spots of the sample, resulting in directly stacked graphene.

Stacked graphene was also measured by X-ray diffraction (Figure S3 in Supporting Information). These two results were mainly attributed to the graphene in the sponge, whereas region 3 was due to aggregates of CNF. Similar to the pure CNF sample (CS-Et1500), a high $I_D/I_G$ ratio and low $I_{G'}/I_G$ ratio were obtained for region 3. The high defect density and strong interlayer interaction were features of graphitized CNF. Improving the dispersion of CNF should realize more uniform formation of bulk-scale graphene with a low interlayer interaction.

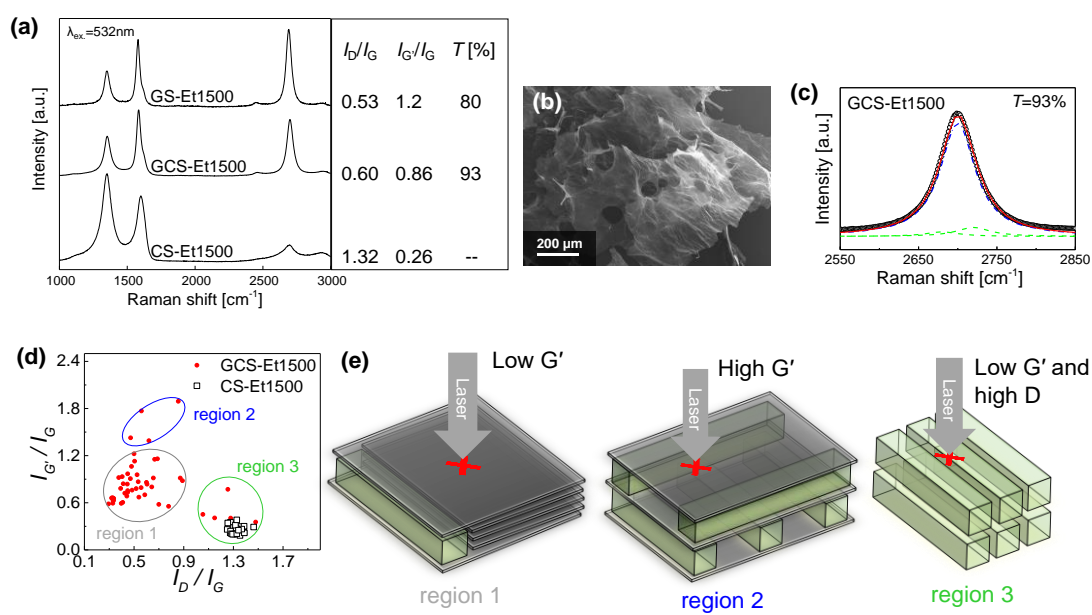

Fig. 5. (a) Raman spectra of GS-Et1500 and GCS-Et1500. Characteristic ratios are indicated to the right of the corresponding spectra. (b) SEM image of GCS-Et1500. (c) G′-band fitting of GCS-Et1500. (d) Distribution of $I_D/I_G$ and $I_{G'}/I_G$ for each measurement spot of the GCS-Et1500 (red dot) and CS-Et1500 (hollow square). (e)

Schematic images of three different situations of GCS-Et1500, indicating stacked graphene (region 1), graphene intercalated with CNF (region 2), and aggregate of CNF (region 3).

**4. Conclusion**

By freeze-drying of a GO dispersion and an ethanol-mediated reduction at ultrahigh temperature, we realized a graphene sponge with a high random-stacking fraction for both surface and internal areas. This feature is in sharp contrast to GO aggregate samples, where the formation of AB stacking cannot be suppressed in the internal area. The high random-stacking fraction in internal regions is attributed to the increased accessible area of the porous graphene sponge for ethanol-derived species. We optimized the reduction conditions and confirmed that the strong and sharp $G'$-band from graphene sponge is reduced at 1500 °C, which is indicative of a weak interlayer interaction. Additionally, CNF was introduced as a spacer into a GO sponge to separate the graphene layers and to avoid direct stacking. Blending of CNF with the GO dispersion further reduces the AB-stacked fraction. Although there is still room to improve mixing to achieve a higher $I_{G'}/I_G$ ratio and homogeneity, the proposed scheme prevents strong interlayer stacking in bulk-scale graphene. Consequently, it should realize the scalable production of high-performance bulk-scale graphene in which the superior properties of single-layer graphene are effectively preserved.

**Acknowledgments**

The authors thank Dr. R. Negishi for the fruitful discussion and technical assistance. We also appreciate Mr. T. Mikazuki and DKS Co. for providing the cellulose nanofiber samples. Part of this work was supported by JSPS KAKENHI (Grant Numbers JP15H05867, JP17H02745, and JP19H04545) and Tanikawa Fund Promotion of Thermal Technology. SEM observation was performed at the Photonics Center, Osaka University.

Supporting information

# Bulk-scale synthesis of randomly stacked graphene with high crystallinity

Zizhao Xu[a], Shingo Nakamura[a], Taiki Inoue[a], Yuta Nishina[b], Yoshihiro Kobayashi[a]

[a] Department of Applied Physics, Graduate School of Engineering, Osaka University, 2-1 Yamadaoka, Suita, Osaka, 565-0871, Japan

[b] Research Core for Interdisciplinary Sciences, Okayama University, 1-1-1 Tsushima-naka, Kita-ku, Okayama, 700-8530, Japan

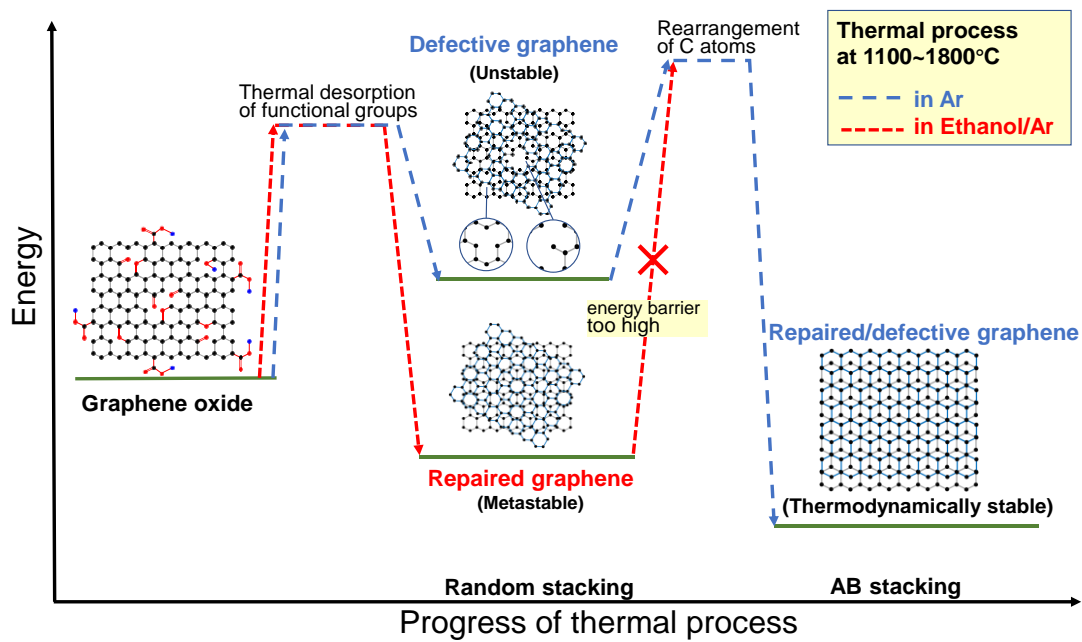

Figure S1 Energy diagram for the formation of graphene with random stacking from graphene oxide by the thermal process.

Graphene oxide (GO) containing various functional groups constructs a random stacking structure in bulk scale instead of ordered AB stacking in graphite and usual multi-layer graphene because of relatively weak interaction between GO flakes. In this work, graphene in bulk scale was synthesized from GO by the thermal process at high temperature. During the high-temperature process, the functional groups of GO were dissociated and desorbed subsequently from the GO surface, leaving randomly oriented graphene with large numbers of vacancy defects and dangling bonds. Considering GO samples treated under only inert gas, the formed graphene with the vacancy and dangling bonds becomes in a metastable state but remains higher energy level. Consequently, the activation energy barrier to the thermodynamically stable AB stacking [1] is relatively low enough to overcome by thermal excitation around 1800 °C, resulting in the structural transformation from random stacking to AB stacking, and the low randomly stacking fraction of GA-Et1800, GA-Ar1800-surf, and GS-Ar1800-surf. On the other hand, in the case of the thermal process with ethanol addition (GS-Et1800 and GA-Et1800-surf), decomposed ethanol provides reactive species containing carbon and oxygen, which effectively repair the vacancies with dangling bonds and topological defects in reduced GO[2, 3]. The reduced defect density causes the formed graphene with random stacking to fall in the metastable state with a very stable energy level. Consequently, the activation energy barrier to the ordered AB stacking becomes too high to get over by thermal process at 1800 °C or below, resulting in the formation of the turbostratic graphene with low defect density. It should be noted that the turbostratic graphene formation will be limited for the thermal process under 2000 °C since the turbostratic graphene with random stacking will also turn into the most stable AB stacking graphene [4] by thermal excitation above 2000 °C as the result of overcoming the activation barrier. It should be also pointed out that the rGO treated in pure Ar (e.g. GS-Ar1800-surf) possesses higher defect density and a higher fraction of AB stacking compared with GO treated in the ethanol environment (e.g. GS-Et1800-surf) according to the results of Raman analysis in Fig. 2 (a). This result means defects remaining after the thermal process at 1800 °C should not hinder the transition to AB-stacking but may assist the process possibly due to lower energy barrier for the rearrangement of the chemical bonds.

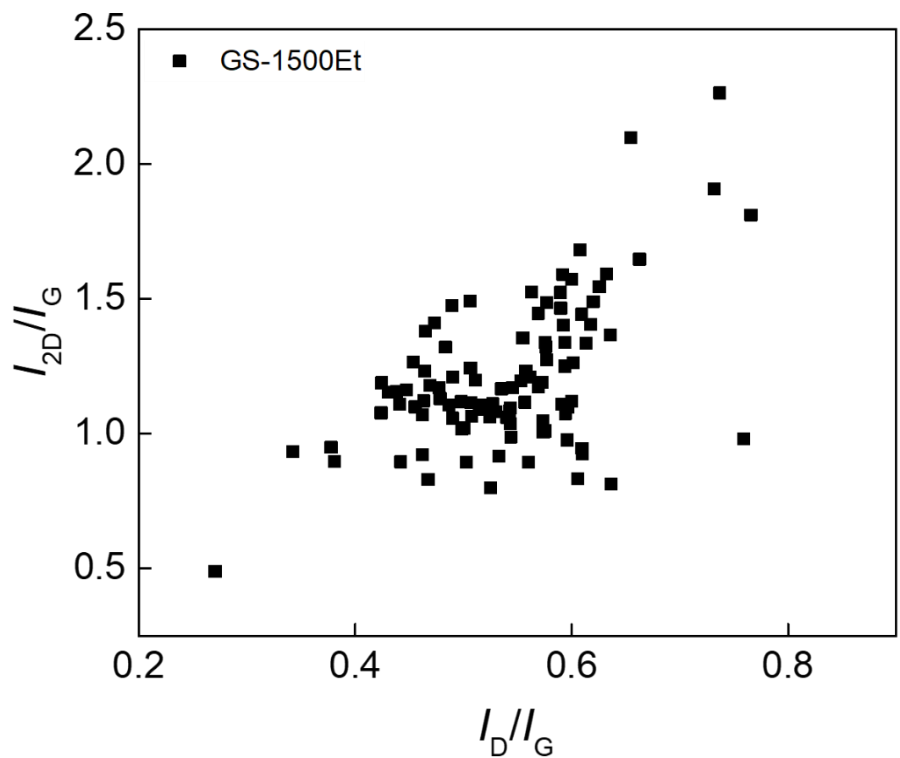

Figure S2 $I_{G'}/I_G$ vs $I_D/I_G$ of GS-Et1500.

The majority of the data points is located within the ranges of 0.9–1.6 for $I_{G'}/I_G$ and 0.45–0.65 for $I_D/I_G$. The standard deviations of $I_{G'}/I_G$ and $I_D/I_G$ are 0.1 and 0.27, respectively. The origin of the variation is attributed to the spatially different morphology of the sponge structure formed during the freeze-drying process. While the porous morphology assists the ethanol-mediated defect healing, locally aggregated graphene in some regions might be less healed. We consider a further enhancement of uniformity is possible by improving the forming process of the sponge structure.

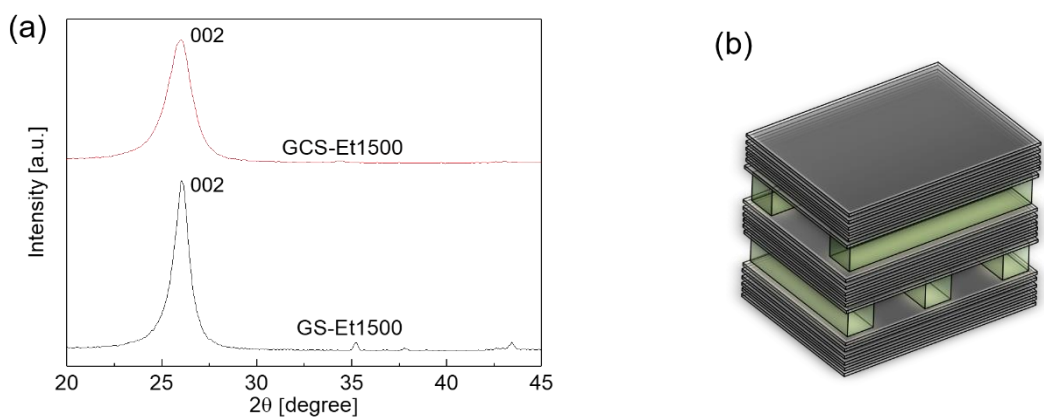

Figure S3 (a) XRD result of GS-Et1500 and GCS-Et1500. (b) Schematic structure of GCS-Et1500.

X-ray diffraction (XRD) measurement was conducted on GS-Et1500 and GCS-Et1500 (Figure S2 (a)). 002 peak from GS-Et1500 and GCS-Et1500 were observed. *K* value of turbostratic graphene was 0.9 on the c-axis[5]. According to Scherrer equation, the crystallite sizes on c-axis of GS-Et1500 and GCS-Et1500 were 11 nm and 9 nm, respectively. The average layer distances of GS-Et1500 and GCS-Et1500 were 0.342 nm and 0.343 nm, which were larger spacing than that of graphite (0.335 nm). GCS-Et1500 showed larger layer distance and smaller crystallite sizes on c-axis, indicating that cellulose nanofiber (CNF) was intercalated between graphene layers. Nanostructure of our sample can be schematically depicted in Figure S3(b), in which graphene layers are not individually separated by CNF but small stacks of graphene are separated by CNF.